\documentclass[aps,prb,twocolumn,superscriptaddress,showpacs,floatfix,amsmath,amssymb,10pt]{revtex4-1}

\usepackage{hyperref}
\usepackage{graphicx}
\usepackage{physics}	

\begin{document}
 
\title{Identifying structural changes with unsupervised machine learning methods}

\author{Nicholas Walker}
\affiliation{Department of Physics \& Astronomy, Louisiana State University, Baton Rouge, Louisiana 70803, USA}
\author{Ka-Ming Tam}
\affiliation{Department of Physics \& Astronomy, Louisiana State University, Baton Rouge, Louisiana 70803, USA}
\affiliation{Center for Computation \& Technology, Louisiana State University, Baton Rouge, Louisiana 70803, USA}
\author{Brian Novak}
\affiliation{Department of Mechanical Engineering, Louisiana State University, Baton Rouge, Louisiana 70803, USA}
\author{M.\ Jarrell}
\affiliation{Department of Physics \& Astronomy, Louisiana State University, Baton Rouge, Louisiana 70803, USA}
\affiliation{Center for Computation \& Technology, Louisiana State University, Baton Rouge, Louisiana 70803, USA}

\begin{abstract}
Unsupervised machine learning methods are used to identify structural changes using the melting point transition in classical molecular dynamics simulations as an example application of the approach. Dimensionality reduction and clustering methods are applied to instantaneous radial distributions of atomic configurations from classical molecular dynamics simulations of metallic systems over a large temperature range. Principal component analysis is used to dramatically reduce the dimensionality of the feature space across the samples using an orthogonal linear transformation that preserves the statistical variance of the data under the condition that the new feature space is linearly independent. From there, k-means clustering is used to partition the samples into solid and liquid phases through a criterion motivated by the geometry of the reduced feature space of the samples, allowing for an estimation of the melting point transition. This pattern criterion is conceptually similar to how humans interpret the data but with far greater throughput, as the shapes of the radial distributions are different for each phase and easily distinguishable by humans. The transition temperature estimates derived from this machine learning approach produce comparable results to other methods on similarly small system sizes. These results show that machine learning approaches can be applied to structural changes in physical systems.
\end{abstract}


\maketitle

\section{Introduction}

Machine learning (ML) has seen rapid development over the last decade or so. At present, rather sophisticated packages are readily available for the application of various ML methods.\cite{scikit-learn,r_package} Conceptually, the ML approach can be regarded as a data analysis approach for detecting patterns in data and then using the extracted patterns for classification or regression. Modern scientific investigations, in particular numerical study, naturally involve large data sets. However, conventional approaches often neglect possible nuance the structure of the data. Although inference methods, such as the maximum likelihood method and the maximum entropy method\cite{PhysRevB.44.6011,JARRELL1996133} have been routinely applied on certain physical problems, applications which utilize other ML methods have not attracted much attention until recently. The advances in ML algorithms and implementations provide a exciting new proposal for applying them to understanding data from physical sciences and perhaps improving upon existing numerical methods.\cite{PhysRevB.95.035105}

In contrast with conventional approaches, ML provides a new avenue for unveiling the underlying structure in data beyond simply measuring the mean, variance, or higher moments of the data. This provides not only a new method for a deeper understanding of old problems, but also new problems which have been hitherto impossible to approach and therefore ignored. An interesting topic which can benefit from this new approach is the study of interacting systems at both quantum and classical scales. Currently outstanding problems include the calculation of phase transition points and the prediction of phase diagrams. Some remarkable recent papers have shown that certain ML algorithms can be used to identify phase transitions of lattice models, particularly on spin systems.\cite{carrasquilla2017machine,PhysRevB.94.195105}

Utilizing the ML approach for studying phase transitions implicitly assumes that there is some form of change in the pattern of the measured data across the phase transition. This is in fact exactly what happens in most phase transitions. For instance, in the melting of a crystal, the widely adopted Lindemann parameter is essentially a measure of the deviations of atomic positions in the system from equilibrium positions.\cite{lindemann} Similar behavior is found in most phase transitions of molecular systems, often in the form of pattern changes in the atomic positions. Perhaps more importantly, for sufficiently complex systems, their phase transitions do not have obvious order parameter or the order parameter is simply an unknown, often prohibiting the detection of such pattern changes. This is not a hypothetical situation, indeed this is the case for some interesting materials, such as heavy fermion materials.\cite{PhysRevLett.96.036405} ML is a new route of studying those transitions by searching for hidden patterns in the measured data. 

Pattern recognition is a strong suit of ML methods and many existing applications of ML methods are designed for identifying patterns in figures, such as the classic example of handwriting recognition.\cite{writing} Given the relation between phase transitions and pattern changes in measured data, using ML methods to identify phase transitions is an attractive prospect. There are two major categories of ML methods, supervised and unsupervised. Both categories have been considered as good candidates for identifying phase transition in the lattice Ising model in two dimensions.\cite{carrasquilla2017machine,PhysRevB.94.195105,lattice_spin,ising_model} For the well studied Ising model, in which the phase diagram and even the critical point are known exactly in two dimensions, \cite{PhysRev.65.117} those known results facilitate the application of supervised ML methods. There have also been efforts in classifying crystal structures and predicting melting points in octet AB solids.\cite{structure_classification} For a system with minimal a priori information, applying a supervised ML method may be challenging. As interest in multi-component high entropy alloys grows, oftentimes a problem arises in the fact that the phase diagrams are largely unknown. For this reason, we decide to explore an unsupervised ML method in this paper.

The pattern recognition capabilities of ML methods can be applied to structural changes found in physical systems by partitioning a large data set of structural information according to a similarity criterion into distinct classes. The distributions of the partitions with respect to some physical property associated with the structural change can be used to predict the transition point. As such, this approach is an empirical method for detecting structural changes.

As an example of the application of this unsupervised machine learning approach to detecting structural changes, this paper focuses on the detection of the solid-liquid phase transition in small titanium and aluminum systems by analyzing the structural information of classical molecular dynamics (MD) simulations about the melting point. Within the context of this computation, the approach described above is similar to the single-phase hysteresis method, which is an existing empirical method for calculating melting points. This involves either heating a bulk solid until it melts or cooling a bulk liquid until it solidifies at fixed pressure, using an order parameter to classify the system as solid or liquid. This method incurs a large error due to the effect of either overheating or undercooling, where the material melts at higher temperatures and solidifies at lower temperatures in MD simulations with respect to the experimental data. The discrepancy can be as large as 20\% at the same pressure \cite{PhysRevB.72.125206,doi:10.1063/1.4794916}. This is largely due to the surfaceless feature of the bulk material, which inherently restricts the nucleation to being homogeneous rather than heterogeneous. This can be assuaged through the use of the hysteresis method, which is empirically based in nucleation theory.\cite{noneq_melt_cryst} The melting temperatures from the heating and cooling methods, $T_M^+$ and $T_M^-$, can respectively be used to establish a melting temperature $T_M$ according the the equation $T_M = T_M^+ + T_M^- - \sqrt{T_M^+ T_M^-}$, which can be in good agreement with experiment.\cite{melting_cu} Due to its status as an empirical method, however, there is no physical significance to the relationship itself. Error analysis from this method is also difficult to quantify. It is also worth mentioning that depending on the system and the experimental conditions, overheating and undercooling effects can also be seen in experiments. The approach presented in this paper can be thought of as an automated version of this empirical method, using unsupervised machine learning to classify the crystal structure instead of an order parameter, which requires a priori information. See Appendix A for information about additional methods for calculating melting temperatures of materials.

The organization of this paper is as follows. In the next section, we explain, in detail, the ML algorithms for the present project. In section III, we present the main results from the calculations for titanium and aluminum. In the last section, we conclude and discuss the future directions for applying ML approaches to molecular systems.

\section{\label{ml_method}Unsupervised machine learning method for calculating melting points}

The machine learning approach used in this paper follows a procedure similar to that used in the single phase method described in the preceding section. A perfect lattice is heated from a temperature well below the melting point to a temperature well beyond the melting point. Then, the system is cooled back down to the original starting temperature. In this manner, both the melting and solidification phenomena are captured. The MD itself should be considered as a sampling method, as we do not study the dynamics explicitly in this study. One of the foremost ways that humans distinguish solid and liquid structures in MD data obtained for crystalline materials is the radial distribution function. In principle, machine learning methods can be applied to achieve a similar result, but with much higher throughput. For this study, the radial distributions for a subset of the simulation steps were passed through a clustering algorithm. Qualitatively, this means that the dataset is partitioned into two groups based on a measure of similarity. 

The notion of what ``similarity" means varies from method to method. For the k-means algorithm, originally from signal processing, $n$ samples are partitioned into $k$ clusters such that the clusters exhibit prototypical centroids by which each sample is grouped with according to geometrical proximity.\cite{kmeans_orig, kmeans_coin, kmeans_algo} The Euclidean metric is used for this application and the procedure iteratively produces the best choice of centroids by restarting multiple times to avoid falling into local minima.\cite{scikit-learn} This is done by making an initial random choice of centroids from the sample space, classifying the remaining samples by proximity to the chosen centroids, then updating the choice of centroids based on the average positions of the member samples in each cluster. The procedure is considered complete when the centroids no longer shift beyond a defined threshold in the update step. In this application, $k=2$ and the two clusters are intended to represent the solid and liquid phases. See Appendix C for detail about the k-means clustering procedure.

The clustering results using the radial distribution data itself is not particularly easy to visualize, as the dimension of the feature space for the samples is the number of bins used for constructing the radial distribution data. The dimensionality of this data can be reduced using principal component analysis (PCA).\cite{pca} This procedure involves performing an orthogonal linear transformation to a new feature space of equal or lesser dimension such that the principal components composing the projections on to the new feature space guarantee the largest explained variance between the samples under the condition that the features are linearly independent. The principal components are ordered by their explained variance ratios. This smaller feature space is easier to analyze and allows for the curse of dimensionality to be avoided while simultaneously ensuring that the statistically significant features of the data are preserved and easy to demonstrate. Prior to performing PCA reduction, it is important to always scale the features beforehand to prevent inappropriate domination of a subset of the features over the others. Min-max scaling was used in this application such that each of the features shared a common domain ranging from 0 to 1. See Appendix B for detail about the PCA procedure. 

After performing a PCA feature space reduction with two principal components, k-means clustering with two clusters is used to partition the structural data into two groups, clusters $A$ and $B$, according on structure similarity.  Note that the chosen clustering method may need to be adjusted depending on the shape of the data produced by the PCA reduction. For example, clustering for data that exhibits irregular boundaries or is well-connected but not necessarily dense may perform better with agglomerative clustering or spectral clustering. These cluster centroids are the prototypical radial distributions that the samples in their respective clusters most closely resemble as a group, at least in terms of minimizing the variance of the samples within the clusters. The cluster labels can also be verified to be the same or similar for both the raw and the PCA reduced data to ensure that the important statistical features are being preserved by the PCA procedure. Once the clusters are verified to represent the desired phases, a temperature distribution is constructed for each cluster. There will be a range of temperatures for which the cluster temperature distributions overlap which will be referred to as the transition region since samples that belong to either cluster coexist within this temperature range. To investigate the transition region more closely, the contributions from each of the cluster distributions that are contained by the range of temperatures defined as the transition region are isolated by truncating the full cluster distributions such that they are restricted to the transition region. The truncated distributions are then renormalized such that correct means and standard deviations can be extracted from them. The mean temperatures of the truncated distributions, $T_A$ and $T_B$, can be averaged in various ways to estimate the transition temperature. In this paper, arithmetic and geometric means were used.

A similar approach using PCA has been recently explored for the two-dimensional lattice Ising model, in which snapshots of the spin configurations are used as learning samples to detect the second order phase transition.\cite{PhysRevB.94.195105} The present work, when applied to the detection of melting points, can be considered as a generalization of this approach, with continuous variables instead of discrete spin, a continuum instead of a lattice, and a first order transition instead of a second order transition. Additionally, we further refine the method used in the Ising model by clustering the data after the PCA reduction and by estimating the transition temperature through analysis of the clustered data distributions.

\section{Results}

This method is applied to two small systems of 128 titanium atoms and 108 aluminum atoms. We choose these two systems as representative examples of metals with body-centered cubic (BCC) and face-centered cubic (FCC) crystal structures. The MD simulations were carried out with the LAMMPS simulation package.\cite{lammps} The titanium potential used is a modified embedded atom model (MEAM) spline potential specifically made for describing phase transitions of titanium with a stable titanium-$\beta$ phase.\cite{ti_meam} The aluminum potential used is also a MEAM potential, albeit not a spline function.\cite{ti_al_meam, ti_al_other_meam} For each MD simulation phase, the systems were held at 0 bar in the isobaric-isothermal (NPT) ensemble using 3-chain Nose-Hoover thermostats and barostats with damping parameters of 128 and 1024 time steps, respectively, and a simulation time step of .00390625 ps. The titanium system is initialized in the BCC titanium-$\beta$ phase with velocities generated by a Gaussian distribution to produce a temperature of 1280K and the aluminum system is initialized in the FCC configuration with velocities generated in the same manner to provide a temperature of 256K. A null value was enforced for the aggregate linear and angular momenta of both systems when generating the velocities. In the first MD phase, the systems are held at the initial temperature for 2.048 ns. The second phase then ramps the temperature to 3072K for the titanium system and 2560K for the aluminum system over 8.192 ns. The systems are then held at those maximum temperatures for 2.048 ns in the third phase before ramping back down to the initial temperature over 8.192 ns in the fourth and final phase. Only the second and fourth phases that respectively characterize the melting and solidification processes are included in the data analysis. The energy, pressure, volume, and atomic position data are recorded every 32 time steps. The datasets generated and/or analyzed during the current study are available from the authors upon reasonable request. In other simulations, various reasonable minimum and maximum temperatures were tested with the method without significant impact on the final results.

After all of the data is collected, the radial distribution functions are calculated with 256 bins for all of the recorded time steps out to $\sqrt{3}l_0/2$ where $l_0$ is the minimum side length of the simulation box, with each value in the function acting as a feature and each function itself acting as a sample for the purposes of the machine learning approach to analysis. All work-flow and post-processing programming are done with Python.\cite{python} For handling large data arrays, the NumPy Python library is used.\cite{numpy} The scikit-learn Python package is used to perform the PCA and k-means procedures.\cite{scikit-learn} All plots were made using the Matplotlib library for Python and the perceptually uniform ``plasma'' color map is used for all color maps.\cite{matplotlib}

\begin{figure}[!htb]
\centering
\includegraphics[width=\columnwidth]{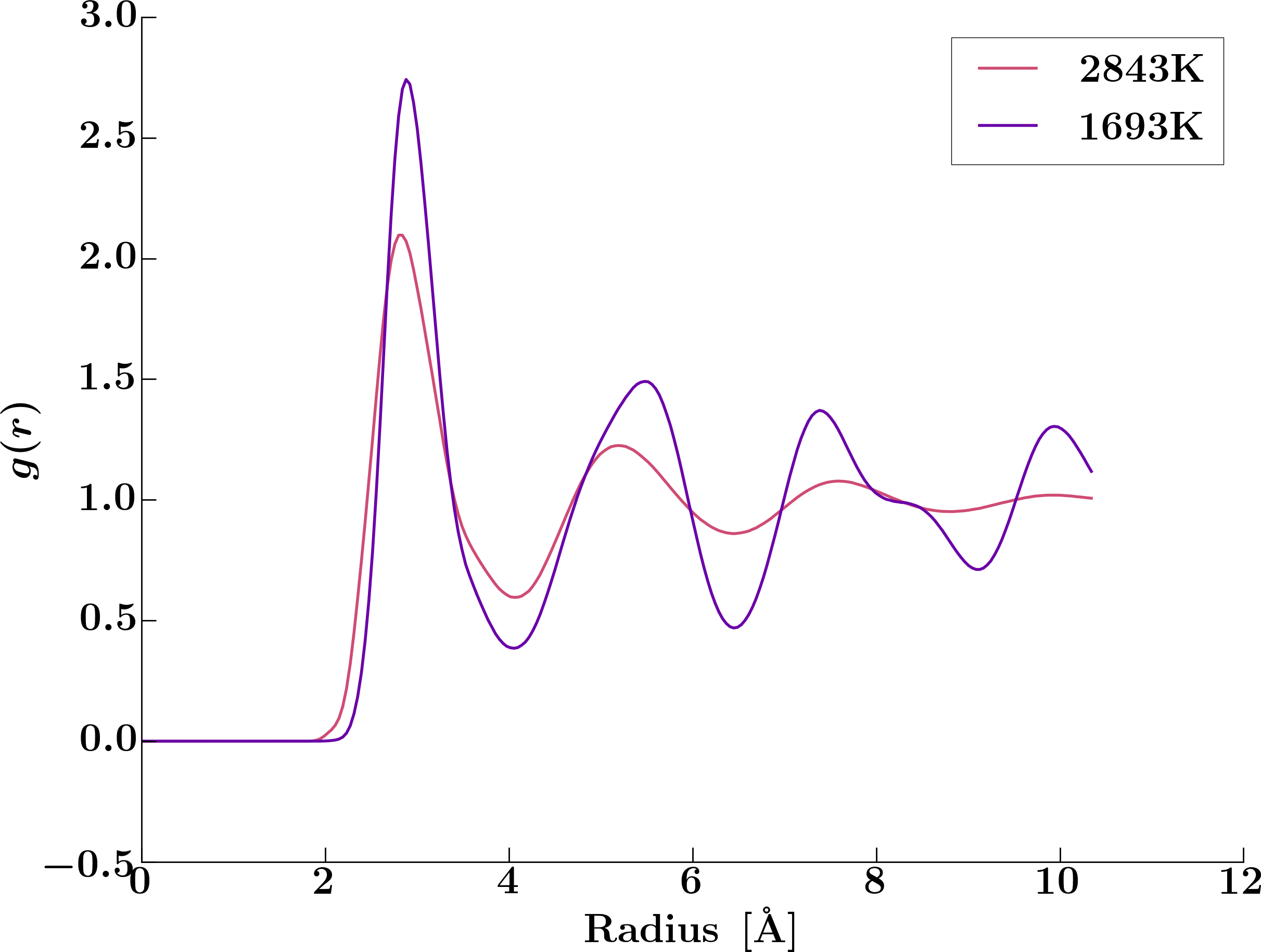}
\caption{K-means cluster centroids for the unreduced titanium radial distribution function data. These cluster centroids represent the prototypical radial distributions for each cluster.}
\end{figure}

\begin{figure}[!htb]
\centering
\includegraphics[width=\columnwidth]{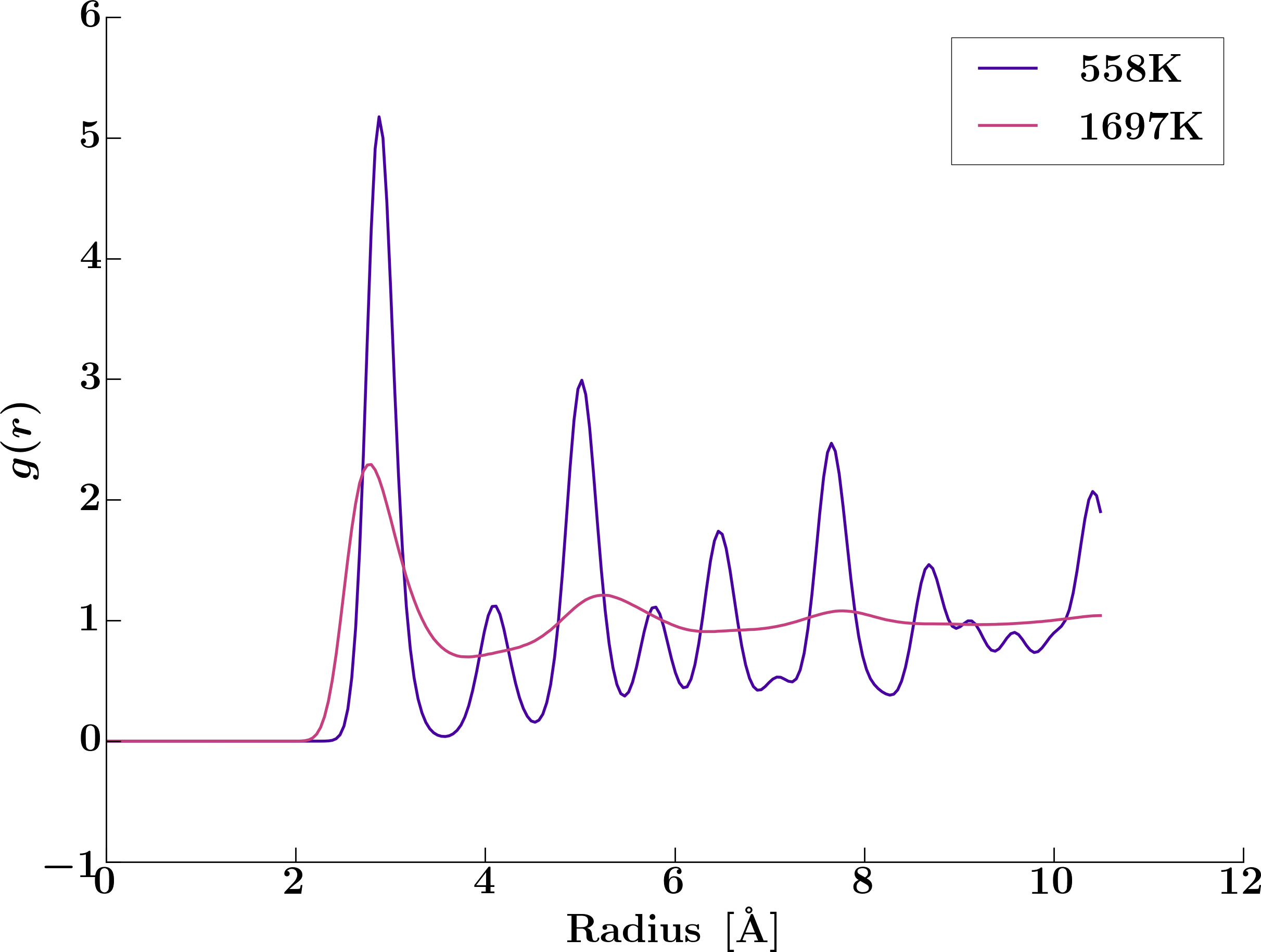}
\caption{K-means cluster centroids for the unreduced aluminum radial distribution function data. These cluster centroids represent the prototypical radial distributions for each cluster.}
\end{figure}

As seen in Fig. 1, the k-means cluster centroids for the unreduced titanium radial distribution data paint a very clear picture of the molecular structure exhibited by the clusters. There are two clusters, one cold ($A$), with an average temperature well below the reported melting point for the potential of around 1900K,\cite{ti_meam} and one hot ($B$), with an average temperature well above the known melting point. The centroid of cluster $A$ very strongly resembles that of an equilibrium BCC lattice, albeit smoothed out a bit. This is consistent with the initial structure of the titanium-$\beta$ system. Cluster $B$ resembles that of a typical liquid, albeit with a slight kink near the first local minimum after the first shell. These results would suggest that the k-means clustering is indeed clustering the samples in a manner that is consistent with the expected structures in the sample space.

Similar to the titanium data, the k-means cluster centroids for the unreduced aluminum radial distribution data in Fig. 2 show the molecular structure exhibited by the clusters. The colder cluster ($A$) has an average temperature well below the melting point of 937K\cite{ti_al_other_meam} and the warmer cluster ($B$) also has an average temperature above the melting point. Cluster $A$ resembles a softened FCC equilibrium lattice, which is consistent with the initial FCC structure as expected. Cluster $B$ also clearly resembles the structure of a liquid with the same small kink in the first minimum after the first shell that was seen in the titanium results. Once again, the k-means clustering results are consistent with the expected partition of the samples by molecular structure into solid and liquid phases.

For both systems, the results of the k-means clustering suggest that the centroids of the clusters can be interpreted as the prototypical solid and liquid structures by which the other samples are categorized with by similarity.

\begin{figure}[!htb]
\centering
\includegraphics[width=\columnwidth]{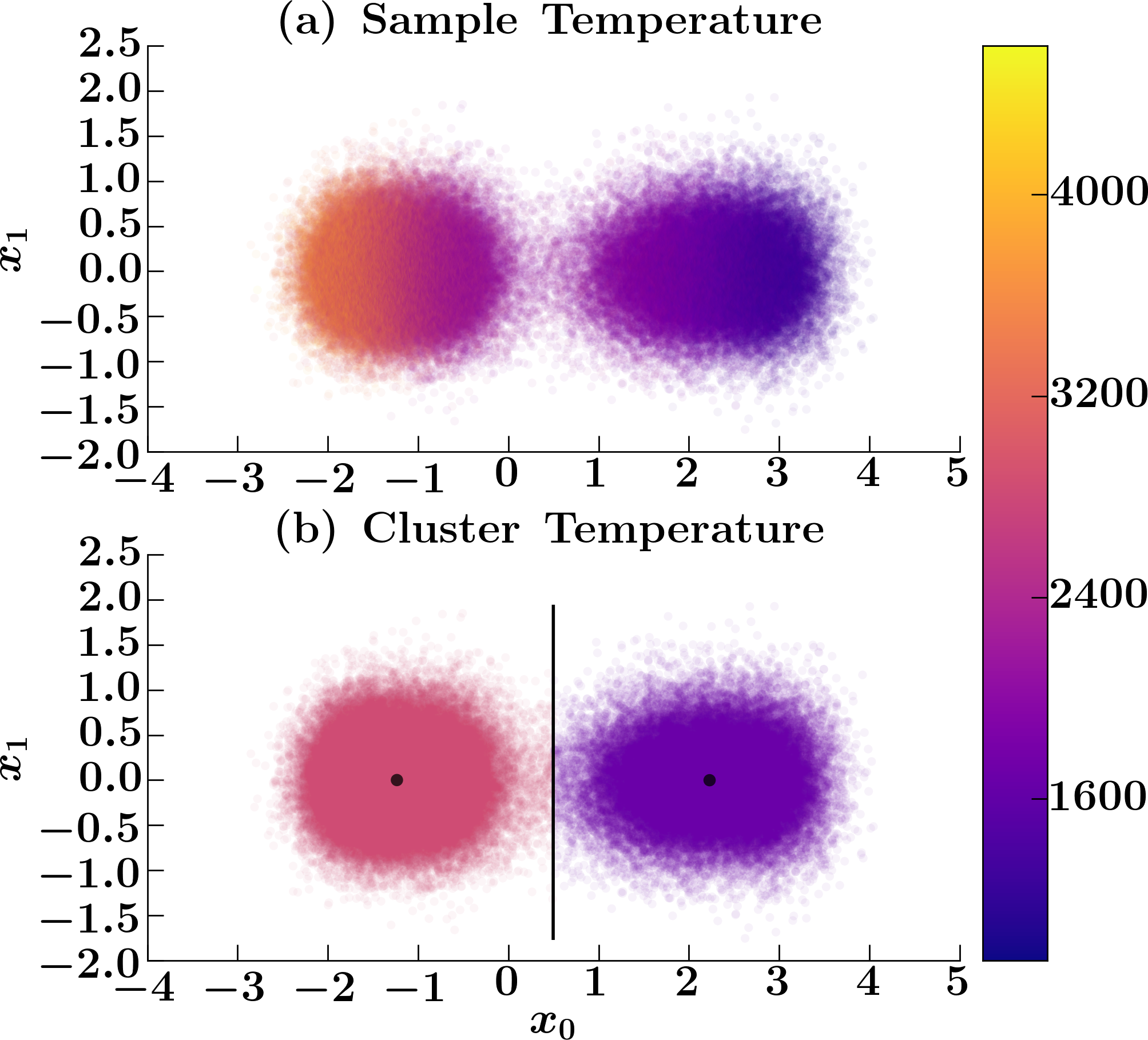}
\caption{Scatter plots of the PCA-reduced two-dimensional radial distribution samples for titanium with respect to the first two principal axes, $x_0$ and $x_1$. (a) Samples colored with respect to their instantaneous temperatures. (b) Samples are colored by cluster temperature average with a line marking the cluster boundary and circles marking the cluster centroids.}
\end{figure}

\begin{figure}[!htb]
\centering
\includegraphics[width=\columnwidth]{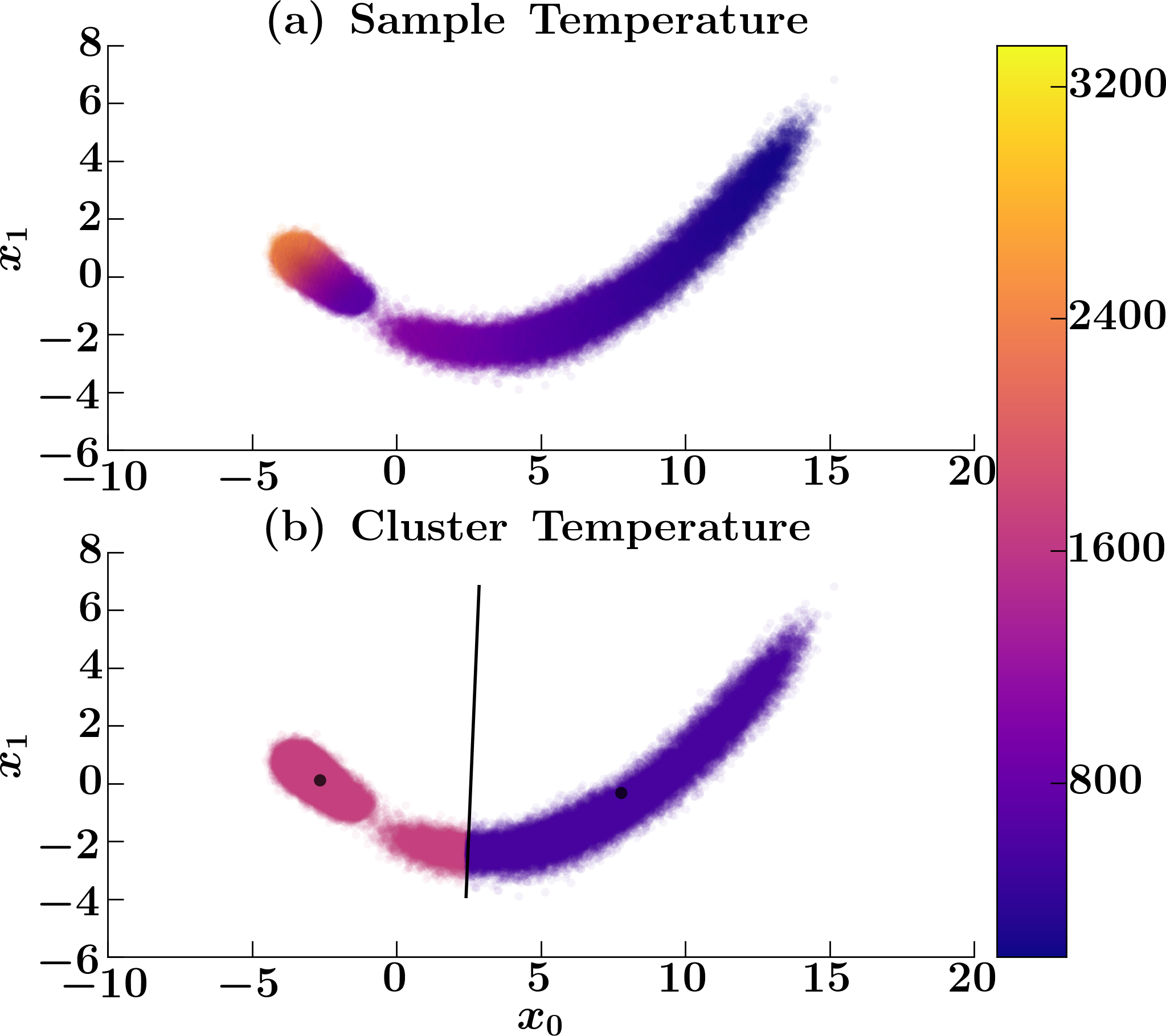}
\caption{Scatter plots of the PCA-reduced two-dimensional radial distribution samples for aluminum with respect to the first two principal axes, $x_0$ and $x_1$. (a) Samples are colored with respect to their instantaneous temperatures. (b) Samples are colored by cluster temperature average with a line marking the cluster boundary and circles marking the cluster centroids.}
\end{figure}

In Fig. 3a, the data shows a very strong relationship between the sample temperature and sample position along the first principal axis ($x_0$) for titanium. The shape of the data also pinches off, almost partitioning the samples along the second principal axis, suggesting a naive clustering. In Fig. 3b, the clustering assignments and the line separating them show that the data is partitioned into clusters slightly to the side of the recess noted in the shape of the data, towards the colder sample. Furthermore, the line is almost vertical with respect to $x_0$, consistent with the temperature gradient along the axis, likely because of of the near-symmetry of the data about the said axis. Indeed, the PCA analysis reports that $x_0$ explains 35.94\% of the variance in the data while the second primary axis ($x_1$) explains only 1.98\% of the variance in the data.

Fig. 4a also shows a very strong relationship between the sample temperature and the sample position along $x_0$ for aluminum. Once again, there is a pinching in the data for aluminum, though not quite along $x_0$ as with titanium. The shape of the aluminum samples is more curved and lacks the symmetry about $x_0$ that the titanium data exhibited. However, the aluminum samples are more stretched along $x_0$ than with the titanium samples such that said axis explains 73.95\% of the variance in the data and $x_1$ explains 4.39\%. The clustering results show that the boundary is not quite vertical, as one could presume from the curvature of the data, and the boundary is once again slightly biased towards samples colder than the location of the recess in the data, albeit a bit more biased than was seen with the titanium data. The reason for the elongated ``tail'' in the data seen in the aluminum samples but not in the titanium samples may be due to the different initial structures of the two metals. An FCC lattice can be expected to exhibit much more variation in structure as it approaches its melting point since it bears much less similarity than a BCC lattice does to a liquid structure.

For both systems, there is ample evidence that the PCA analysis is adequately capturing the statistical significance of the features and that a third principal axis is unneeded as it necessarily explains less variance than the axis that precedes it. Furthermore, the cluster assignments are found to be nearly identical to the unreduced case, with some minor deviations near the cluster boundary for both systems. Thus, PCA reduction is found to be useful for representing the data in a much more easily visualized manner with very minor loss of statistical significance or fidelity in the analysis for both systems. 

\begin{figure}[!htbp]
\centering
\includegraphics[width=\columnwidth]{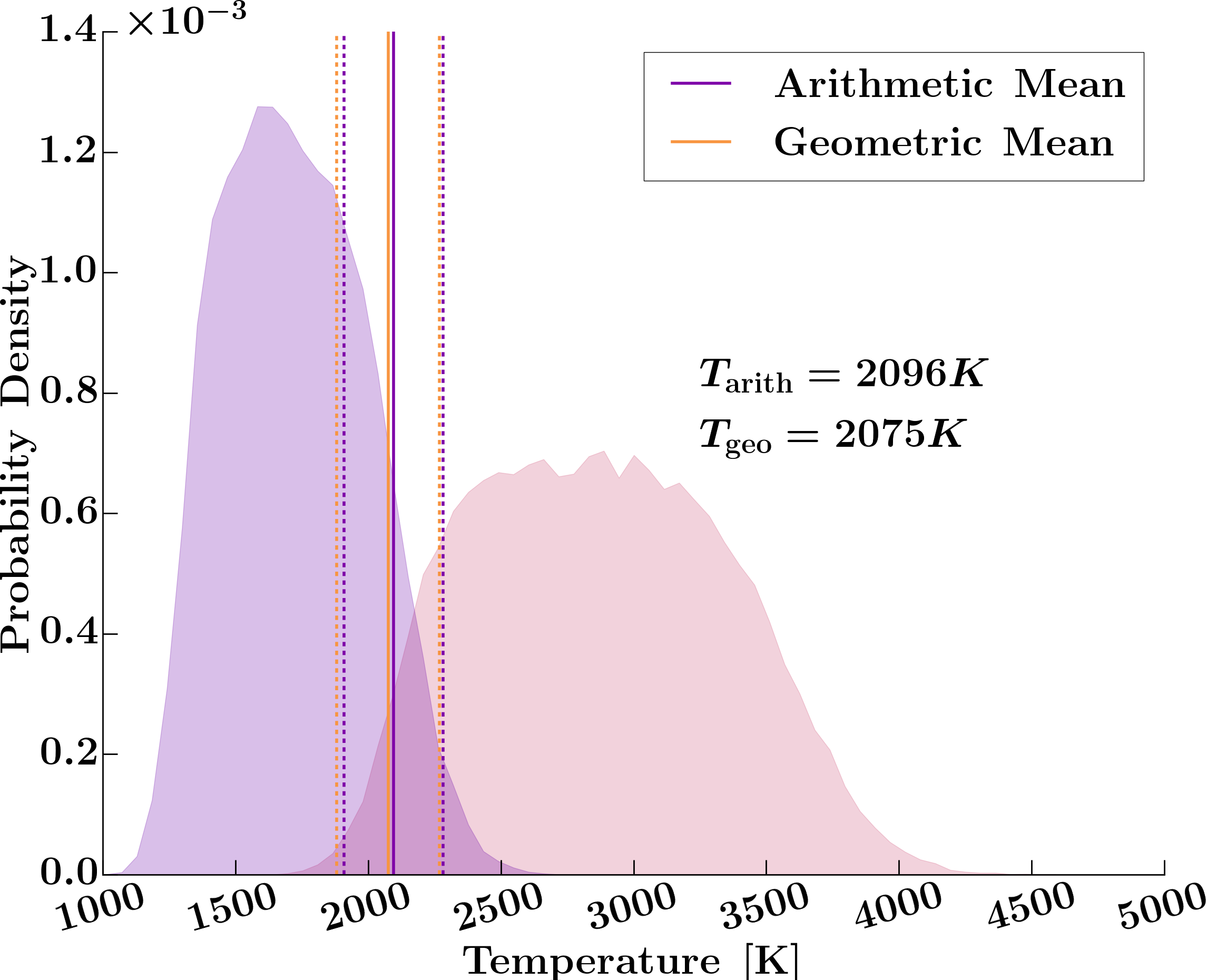}
\caption{Temperature distributions for the PCA reduced titanium radial distribution clusters with arithmetic and geometric mean temperatures of the overlap region truncated distributions.}
\end{figure}

\begin{figure}[!htbp]
\centering
\includegraphics[width=\columnwidth]{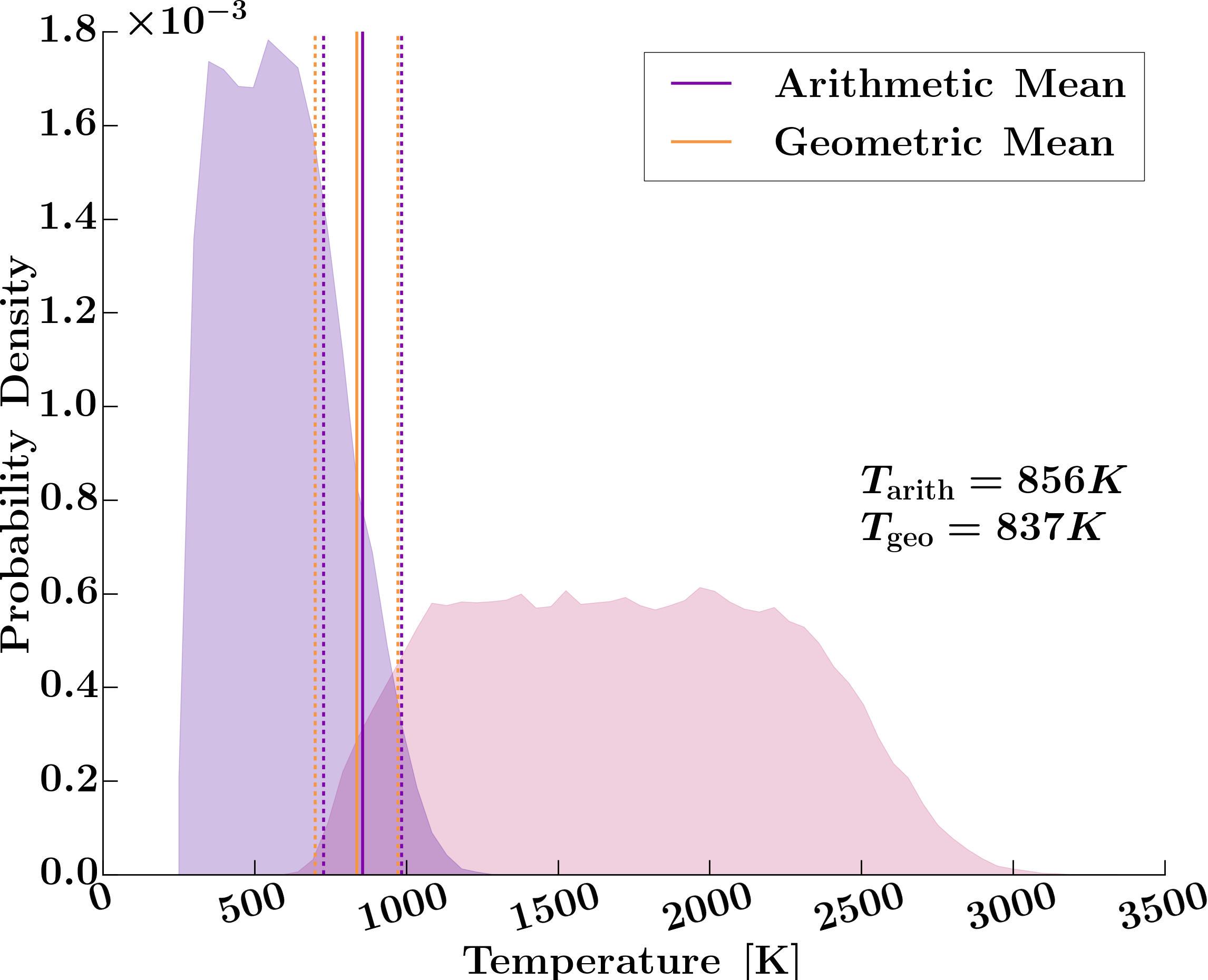}
\caption{Temperature distributions for the PCA reduced aluminum radial distribution clusters with arithmetic and geometric mean temperatures of the overlap region truncated distributions.}
\end{figure}

The temperature distributions of the clusters in Figs. 5 and 6 provide some more detailed insight into the relationship between the temperatures and the structures of titanium and aluminum that are suggested by the PCA reduced scatter plots in Figs. 3 and 4. The overlap region for titanium extends from 1639K to 2604K and the overlap region for aluminum extends from 593K to 1182K. In order to determine the transition temperature between the two structures, the distributions are truncated and renormalized such that the domain of the new distributions lays solely within the overlap region. The mean temperatures of these regions are then determined both arithmetically and geometrically. The arithmetic formulae for determining the means ($T_{\text{arith}}$) and standard deviations ($\sigma_{\text{arith}}$) of each cluster with $N$ samples of temperature $T_i$ and weight $w_i$ (the proportion of the data at temperature $T_i$) within the transition region is as follows

\begin{align}
T_{\text{arith}} &= \sum_{i=1}^N w_i T_i, \\
\quad \sigma_{\text{arith}} &= \sqrt{\sum_{i=1}^N w_i \qty(T_i - T_{\text{arith}})^2}
\end{align}

The corresponding geometric formulae for $T_{\text{geo}}$ and $\sigma_{\text{geo}}$ are then

\begin{align}
T_{\text{geo}} &= \exp{\sum_{i=1}^N w_i \log T_i}, \\
\sigma_{\text{geo}} &= \qty(\exp{\sqrt{\sum_{i=1}^N w_i \log^2\qty(\frac{T_i}{T_{\text{geo}}})}}-1)T_{\text{geo}}
\end{align}

For titanium, the arithmetic mean temperatures of clusters $A$ and $B$ within the overlap region are 1867K and 2325K, respectively, with standard deviations of 175K and 198K, respectively. The arithmetic mean of these values gives a temperature of 2096K with a standard deviation of 186K. The geometric mean temperatures of clusters $A$ and $B$ in the overlap region are 1859K and 2316K, respectively, with geometric standard deviations of 178K and 211K, respectively. The geometric mean of these values then gives a temperature of 2075K with a standard deviation of 194K. The single standard deviation intervals for the predicted transition temperature with arithmetic and geometric averaging are then [1910K, 2282K] and [1881K, 2269K], respectively. The known transition temperature of 1900K is only contained by the interval obtained with geometric averaging.

For aluminum, the arithmetic mean temperatures of clusters $A$ and $B$ within the overlap region are 724K and 989K, respectively, with standard deviations of 115K and 140K, respectively. The arithmetic mean of these values gives a temperature of 856K with a standard deviation of 128K. The geometric mean temperatures of clusters $A$ and $B$ in the overlap region are 715K and 978K, respectively, with geometric standard deviations of 117K and 157K, respectively. The geometric mean of these values gives a temperature of 836K with a standard deviation of 136K. The single standard deviation intervals for the predicted transition temperature with arithmetic and geometric averaging are then [728K, 984K] and [700K, 972K], respectively. The known transition temperature of 937K is contained by both intervals.

\section{Conclusions}

The results show that unsupervised machine learning methods can be used to both isolate the statistically relevant data in structure information as well as cluster that data into groups that represent the expected phases present in a structural change as evidenced by the examples identifying the melting point transitions in titanium and aluminum. The results of the clustering with or without dimensionality reduction are almost identical, indicating that large data sets can be reduced safely without loss of fidelity. The results of the dimensionality reduction with PCA also showed that the majority of the variance in the data is captured by the first principal axis, especially by comparison with the second principal axis. The location of a sample along the first principle axis is also very well-correlated with the temperature of the sample and the partitions made by the k-means clustering are almost exactly along said axis. These features of the unsupervised ML approach to detect structural changes show that this method is promising for unsupervised classification of structures in physical systems.

In the melting point examples, the reported estimates for the melting temperatures are close to but not exactly coincident with the best estimates, though considering the system size, this is to be expected. The melting temperature estimates tended to be off by about 10\% for both cases (9.21\% and 10.67\% for titanium and aluminum, respectively) using the geometric averaging scheme, though the result for titanium is an overestimate and the result for aluminum is an underestimate. The ranges constructed using the standard deviations of the temperatures did indeed capture the true melting temperatures of the potentials in both cases, though the ranges are rather large. Various maximum temperatures well beyond the expected melting point are used to generate MD simulation data and they did not affect the results with any statistical significance. These results are omitted because they provided superfluous information, but serve to indicate that this method is not necessarily sensitive to the temperature range chosen for the simulation data. As it stands, this method does not necessarily outperform existing methods for calculating the melting temperature in terms of accuracy. It may seem to reduce the importance of this approach at first sight, but we stress that these results provide proof of principle for the application of the ML approach to investigating structural changes in physical systems. It has the benefits of conceptual simplicity, ease of application, and speed. More importantly, it can be applied to complicated systems in which there is little a priori data available, contrary to many conventional approaches. 

Further improvements on the method may increase performance, such as improved sampling methods for procuring data, more appropriate choices of structural data, and more complex ML methods. For instance, in the melting point example, Monte Carlo sampling may prove more appropriate than MD sampling\cite{doi:10.1063/1.4794916} as well as a different choice of structural information other than the radial distribution function. It is also worth mentioning that a larger system size can also improve results. For the ML analysis, alternative data scaling, nonlinear dimensionality reduction methods, different clustering methods, or a supervised approach may prove to be more effective.

\section{Acknowledgments}
We thank Shoutian Sun for valuable conversations. This work is funded by the NSF EPSCoR CIMM project under award OIA-1541079. An award of computer time was provided by the INCITE program. This research also used resources of the Oak Ridge Leadership Computing Facility, which is a DOE Office of Science User Facility supported under Contract DE-AC05-00OR22725.

\appendix

\section{Additional Methods for Identifying Melting Points}

Melting is an important physical phenomenon that has inspired the development of many MD methods for investigating said phenomenon. In general, there are three methods that have been used to determine the melting curves of materials using MD that will be discussed here in addition to the single-phase hysteresis method in the main body.

The first method is the two-phase coexistence method which involves allowing a coexisting system to evolve to the temperature at which the free energies of the solid and liquid phases are identical.\cite{coex,asadi_phase-field_2014} This works by providing an initial guess for the melting temperature. A large solid system is then equilibrated at that temperature. Half of the atoms in the system are then frozen while the atoms in the other half are heated to a temperature sufficient for melting. After cooling or rescaling the velocities of the liquid atoms back to the guessed melting point, a short NPT equilibration step is run. If the entire system solidifies or melts during this equilibration run, then the guess is too far from the melting temperature and must be revised. After the NPT run, the system is allowed to evolve for a much longer time in either the microcanonical (NVE) ensemble or more accurately the isobaric-isenthalpic (NPH). During this last step, the system will melt or solidify while the temperature decreases or increases towards the melting temperature due to the latent heat of fusion. From the latter part of the last step, the average temperature is calculated and used as a new guess for the melting temperature. After multiple iterations, the guess will have converged to within uncertainty to the melting temperature; the interface positions and temperature in the final step will be stable and the free energies of the phases will be equal. The physical significance of this method is well-established and there are no superheating/supercooling issues as with the hysteresis method, but to obtain an accurate melting temperature this method requires tens of thousands of atoms and rather long simulation times, which makes the method much more computationally expensive and inaccessible to ab initio MD. However, there are methods for using the coexistence approach with small systems that can predict the melting point within 100K in systems of more than 100 atoms that can be used with ab initio MD.\cite{small_coex}

The second method is the free energy method, which as the name implies involves directly calculating and comparing the free energy from the solid and liquid phases separately. However, for this method, the free energies of solid and liquid phases are calculated by way of thermodynamic integration \cite{free_energy, free_energy_md, free_energy_aimd} from phases with exactly known free energies and with carefully chosen paths to avoid singularity. The melting temperature is obtained when the difference between the free energies is null and the results are generally consistent with the coexistence method.\cite{fe_abinit, fe_coex} This method does not require particularly large systems as it does not require stabilization of the two phases simultaneously, so it can be used for both classical \cite{free_energy_md} and ab initio \cite{free_energy_aimd} MD. However, the thermodynamic integration procedure has to be done with utmost care and for some complicated systems, the reference crystal phase free energy is unknown.  

Lastly, the Z method starts with a perfect lattice that is allowed to evolve in the NVE ensemble. In principle, there is a maximum energy $E_{LS}$ that can be granted to a crystalline system before the system melts.\cite{z_method} If the energy surpasses this quantity, then the system will spontaneously melt at temperature $T_{LS}$, but due to the increase in potential energy largely due to the latent heat of fusion, the temperature will then decrease.\cite{z_method,crit_superheat} The temperature reached during this step coincides with the melting temperature $T_M$. The name of this method is due to the `Z' shape that is traced out by the temperature as a function of the energy as the temperature rises with the energy, suddenly drops at the melting point, then continues to increase. This method can be used with ab initio MD.\cite{z_method_abinit}

However, there are two problems with the Z method involving the waiting time and critical assumptions about melting.\cite{modified_z} First, before the system melts into a liquid, it will stay in the solid phase for a short time called the waiting time that is proportional to both the inverse square of the overheating excess and the inverse of the number of atoms.\cite{z_method_problems} Consequently, the melting results are dependent on the simulation time such that the simulation time must be greater than the waiting time. The second problem is that there is an assumption in the Z method that melting occurs homogeneously throughout the system, which is not true in general, especially as system size and simulation time are increased.\cite{z_method_problems} These two problems are in conflict with one another, but the modified Z method was developed to solve both problems simultaneously by using a parallel piped simulation box such that one dimension is effectively infinite compared to the others and seeking a time evolution into the steady solid-liquid coexistence state.\cite{modified_z}

\section{Principal Component Analysis}

Principal component analysis (PCA) is a linear dimensional reduction algorithm using singular value decomposition to project data into a lower dimensional space.\cite{scikit-learn} The name is in reference to being an analogue of the principal axis theorem from classical mechanics. Assuming that the intial data is encoded in a matrix $\vb{X}$ of shape ($m$, $n$) such that there are $m$ observations for $n$ samples that may be correlated with one another. The goal of PCA is to perform an orthogonal transformation into a new basis set of linearly uncorrelated observations called principal components such that the first one encompasses the largest possible variance in the data and each subsequent principal component also has the largest possible variance under the constraint that they are orthogonal to every preceeding principal component.\cite{pca} Thus, the principal components are guaranteed to be an uncorrelated orthogonal basis set. The dimensionality reduction is accomplished by only considered the first $k$ principal components necessary to capture the variance in the original data set sufficiently well. The mathematical procedure is as follows. The initial data $\vb{X}$ is of the structure

\begin{equation}
\vb{X} = \mqty[\vb{x}_{1} & \dots & \vb{x}_{n}];\quad \vb{x}_i = \mqty[x_{i1} \\ \vdots \\ x_{in}]
\end{equation}

Where each $\vb{x}_i$ contains all of the observations for a sample. Note that at this point, it is assumed that the data is in the mean deviation form such that the mean value for each observation across all of the samples has been subtracted off each entry. The basis of this data is the $m$-dimensional identity matrix. In order to change the basis of the data, consider the linear transformation $\vb{P}\vb{X} = \vb{Y}$ where the new matrix $\vb{Y}$ is the projection of the data for the samples in $\vb{X}$ onto a new basis encompassed by the rows of the matrix $\vb{P}$. This is clear when you write the transformation explicitly.

\begin{equation}
\vb{Y} = \vb{P}\vb{X} = \mqty[\vb{p}_1\cdot\vb{x}_1 & \dots & \vb{p}_1\cdot\vb{x}_n \\ \vdots & \ddots & \vdots \\ \vb{p}_m\cdot\vb{x}_1 & \dots & \vb{p}_m\cdot\vb{x}_n]
\end{equation}

The rows of $\vb{P}$ are defined to be orthonormal such that $\vb{p}_i\cdot\vb{p}_j = \delta_{ij}$. Where $\delta_{ij}$ is the Kronecker delta function. The covariance matrix of $\vb{X}$ is defined as

\begin{equation}
\vb{S}_{\vb{X}} = \frac{\vb{X}\vb{X}^T}{n-1}
\end{equation}

The diagonal entries are the variances of the observations and the off diagonal entries are the covariances of the observations. Thus, the covariance matrix describes the pairwise correlations between all observations. The goal is to determine the basis set $\vb{P}$ such that the off-diagonal elements of the covariance matrix for $\vb{Y}$ ($\vb{S}_{\vb{Y}}$) are minimized. This effectively removes redundancy in the observations and requires the diagonalization of $\vb{S}_{\vb{Y}}$. The covariance matrix of $\vb{Y}$ can be expressed as follows.

\begin{equation}
\vb{S}_{\vb{Y}} = \frac{\vb{Y}\vb{Y}^T}{n-1} = \frac{\qty(\vb{P}\vb{X})\qty(\vb{P}\vb{X})^T}{n-1} = \frac{\vb{P}\vb{X}\vb{X}^T\vb{P}^T}{n-1}
\end{equation}

The matrix $\vb{X}\vb{X}^T$ can be diagonalized such that it is equivalent to $\vb{E}\vb{D}\vb{E}^T$ where the rows of $\vb{E}$ are the right eigenvectors and $\vb{D}$ is a diagonal matrix of the corresponding eigenvalues. The choice of $\vb{P} = \vb{E}^T$ alongside the property of orthonormal matrices such as $\vb{P}$ that $\vb{P}=\vb{P}^{-1}$ gives the result

\begin{equation}
\vb{S}_{\vb{Y}} = \frac{\vb{P}\qty(\vb{P}^{-1}\vb{D}\vb{P})\vb{P}^{-1}}{n-1} = \frac{\vb{D}}{n-1}
\end{equation}

This gives the intended result of minimizing the covariances in the data through an orthogonal transformation. Note that according to the earlier definition of the covariance matrix of $\vb{X}$, the following is true.

\begin{equation}
\vb{S}_{\vb{X}} = \frac{\vb{X}\vb{X}^T}{n-1} = \frac{\vb{P}^{-1}\vb{D}\vb{P}}{n-1}
\end{equation}

Thus, an eigenvalue decomposition of the covariance matrix of $\vb{X}$ must be performed to determine $\vb{P}$, where the rows of the matrix are the principal components. They are ordered form least to greatest variance, which is given by their corresponding eigenvalues. The eigenvalue decomposition is calculated using the singular value decomposition algorithm.

In practical terms, PCA is a useful tool for summarizing data. As was stated prior, the approach is fundamentally intended for reducing redundancy in data. When considering data composed of many observations, one will often find that the observations overlap greatly in common properties amongst the samples that they describe. A naive approach would involve selecting individual observations that appear to contain the most underlying information in the data and ignoring the rest. This risks missing important patterns from the data set by neglecting the possibility of linear combinations of the observations, however. PCA assuages this situation by providing a new orthogonal basis set that maximizes the variance with a linear transformation. The primary weakness of this method is that it can fail to account for nonlinear characteristics in the original data.

\section{K-means Clustering}

K-means clustering is a method of partitioning scattered data into distinct groups. \cite{kmeans_orig, kmeans_coin, kmeans_algo}  Assume that the initial data of the $i$-th data point $\vb{X}$ is an $n-$dimensional vector.

\begin{equation}
\vb{x}_i = \mqty[x_{i1} \\ \vdots \\ x_{in}]
\end{equation}

There are $N$ data points in total. The goal of a general clustering method is to partition these $N$ data points into $k$ different groups according to some criteria. For k-means clustering, the each partition is characterized by the ``center of mass" or centroid, $\vb{C}_{i}$, which is an $n$-dimensional vector. There are a total of $k$ ``center of masses", one for each group that the $N$ data points are partitioned into. The criterion for choosing the centroids and the assignment of a group label to each data point is given by minimizing the total ``moment of inertia",

\begin{equation}
I = \sum_{j=1}^{k} \sum_{\vb{x}_{i} \in \vb{C}_{j}} |\vb{x}_{i}-\vb{C}_{j}|.
\end{equation}

The term $| \vb{x}_{i} - \vb{C}_{j} |$ is defined as the the norm that can be any metric in general. In this work we defined it as $\sum_{k} (x_{ik} - C_{jk})^{2}$. The algorithm used employs an iterative approach to find the centroids commonly referred to as the expectation maximization method.\cite{scikit-learn} This involves initially choosing the centroids, assigning the data points to the centroids by the minimization criterion, then updating the centroids according to the expected mean of the cluster assignments.

\bibliography{ref}
\bibliographystyle{apsrev}
 
\end{document}